\documentclass[runningheads]{lncse}

\usepackage{amsmath}

\begin{document}

\mainmatter   

\title{{\hfill \normalsize BU/HEPP/99-03}\\
Noise Methods for Flavor Singlet Quantities
\thanks{
Talk given at the Interdisciplinary Lattice 
Workshop, Wuppertal, Aug.22-24.}
}

\titlerunning{Noise Methods}

\author{Walter Wilcox}
\authorrunning{Wilcox} 
\institute{Physics Department, Baylor University, Waco TX, USA 76798}

\maketitle   

\begin{abstract}
A discussion of methods for reducing the noise variance
of flavor singlet quantities 
(\lq\lq disconnected diagrams\rq\rq) in lattice QCD is given.
After an introduction, the possible advantage of partitioning
the Wilson fermion matrix into disjoint spaces
is discussed and a numerical comparison of the variance
for three possible partitioning schemes is carried out.
The measurement efficiency of lattice operators is examined
and shown to be strongly influenced by the Dirac and color partitioning
choices. Next, the numerical effects of an automated
subtraction algorithm on the noise variance of
various disconnected loop matrix elements are examined.
It is found that there is a dramatic reduction in the variance of
the Wilson point-split electromagnetic currents
and that this reduction persists at small quark mass.
\end{abstract}

\section{Introduction}

\subsection{Motivations}

The calculation of flavor singlet quantities, also referred to as 
disconnected diagrams because the fermion lines are disjoint, is 
one of the greatest technical challenges left in 
lattice QCD. Disconnected contributions are present in a wide 
variety of quantities in strong interaction physics including 
all baryon form factors, axial operators involving quark spin 
context (Ellis-Jaffe sum rule), hadronic coupling constants 
and polarizabilities, the p-N sigma term, and propagation 
functions for various flavor singlet mesons. Such quantities 
are also present in deep inelastic structure functions measured 
on the lattice using the operator product expansion, but are 
not included because of their difficulty and large Monte Carlo 
error bars. These types of diagrams are difficult to 
evaluate because exact extractions require many matrix 
inversions to measure all the background fermion degrees of freedom 
(including space-time). Disconnected quark 
contributions are instead isolated stochastically by a process of 
applying \lq\lq noises" to the fermion matrix to project out the 
desired operator contribution. 
For an overview of selected aspects of flavor singlet 
calculations in lattice QCD, see Ref.\cite{one}.

\subsection{Mathematical Background}
Noise methods are based upon 
projection of the signal using random noise vectors as input. 
That is, given

\begin{eqnarray}
Mx = \eta, \nonumber
\end{eqnarray}
where $M$ is the quark matrix, $x$ is the 
solution vector and 
$\eta$ is the noise vector, with
\begin{eqnarray}
<\eta_{i}>=0, <\eta_{i}\eta_{j}>=\delta_{ij}, \nonumber
\end{eqnarray}
where one is averaging over the noise vectors, 
any inverse matrix element,
$M^{-1}_{ij}$, can then be obtained from
\begin{eqnarray}
<\eta_{j}x_{i}>= \sum_{k}M^{-1}_{ik}<\eta_{j}\eta_{k}>=M^{-1}_{ij}.
\nonumber
\end{eqnarray}
 
We shall consider two techniques for 
reducing the noise variance in lattice QCD simulations:
partitioning\cite{me1} and subtraction methods\cite{me2}. 
Partitioning the noise appropriately, which means
most generally a zeroing out of some pattern of noise vector elements,
but which will specifically be implemented here by using
separate noise source vectors in Dirac 
and color spaces, can lead to significant reductions in the
variance. Subtraction methods, which involve forming
new matrix operators which have a smaller variance but the
same expectation value (i.e. are \lq\lq unbiased"), can also be of 
great help. The key is using a perturbative expansion of the
quark matrix as the subtraction matrices which, however,
are not unbiased in general and require a separate calculation, 
either analytical or numerical, to remove the bias.
We will see that the various lattice 
operators have dramatically different behaviors
under identical partitioning or subtraction treatments.
Both of these methods, partitioning and subtraction,
will be treated numerically but the hope
is that the numerical results will eventually be
\lq\lq explained\rq\rq by some simple rules 
based on the structure of the Wilson matrix.

\section{Noise Theory}

\subsection{Variance Evaluations}

Let us review the basics of matrix inversion using noise theory.
The theoretical expressions for the expectation value and
variance (V) of matrices with various noises are given
in Ref.\cite{thron}. One has that
\begin{equation}
X_{mn} \equiv \frac{1}{L} \sum_{l=1}^{L}\eta_{ml}\eta^{*}_{nl}.
 \label{element}
\end{equation}
($m,n=1,\dots ,N$; $l=1,\dots ,L$.) We have
\begin{equation}
X_{mn} =X_{nm}^{*},
\end{equation}
and the expectation value,
\begin{equation}
<X_{mn}>=\delta_{mn}.
\end{equation}
By definition the variance is given by
\begin{eqnarray}
V[Tr\{QX\}] \equiv <|\sum_{m,n}q_{mn}X_{nm}-Tr\{Q \}|^{2}>.
\end{eqnarray}
The variance may be evaluated as,
\begin{eqnarray}
&V& [Tr\{QX\}] = \sum_{m\ne n} (<|X_{nm}|^{2}>|q_{mn}|^{2} \\ \nonumber
&+& q_{mn}q_{nm}^{*}<(X_{mn})^{2}> ) 
+ \sum_{n}<|X_{nn}-1|^{2}>|q_{nn}|^{2}.
\end{eqnarray}

\subsection{Real Noises}

For a general real noise,
\begin{eqnarray}
<|X_{mn}|^{2}>=\frac{1}{L},
\label{first} \\
<(X_{mn})^{2}>=\frac{1}{L},
\label{second}
\end{eqnarray}
for $m\ne n$ so that 
\begin{eqnarray}
V [Tr\{QX_{{\rm real}}\}] &=& \frac{1}{L}\sum_{m\ne n} (|q_{mn}|^{2}
+ q_{mn}q_{nm}^{*}) \\ \nonumber
&+& \sum_{n} <|X_{nn}-1|^{2}>|q_{nn}|^{2}.
\end{eqnarray}

The case of real $Z(2)$ has Eqs.(\ref{first}) and (\ref{second})
holding for $m\ne n$, but also
\begin{eqnarray}
<|X_{nn}-1|^{2}>=0.
\end{eqnarray}
This shows that
\begin{eqnarray}
V [Tr\{QX_{Z(2)}\}] \le V [Tr\{QX_{{\rm real}}\}]. 
\end{eqnarray}
Thus, $Z(2)$ noise has the lowest variance of 
any real noise.

\subsection{General Z(N) Noise}

For general $Z(N)$ ($N\ge 3$) noise we have a different situation. One
has that
\begin{eqnarray}
<|X_{mn}|^{2}>=\frac{1}{L},\\
<(X_{mn})^{2}>=0,
\label{ZN}
\end{eqnarray}
for $m\ne n$, and again
\begin{equation}
<|X_{nn}-1|^{2}>=0.
\end{equation}
Thus
\begin{eqnarray}
V [Tr\{QX_{Z(N)}\}] = \frac{1}{L}\sum_{m\ne n} |q_{mn}|^{2},
\end{eqnarray}
and the variance relationship of $Z(2)$ and $Z(N)$ 
is not fixed for a general matrix $Q$. The reason for the
difference in Eqs.(\ref{second}) and (\ref{ZN}) is that the square
of an equally weighted distribution of $Z(2)$ elements is
not itself uniformly distributed (always 1), whereas the square
of a uniformly weighted $Z(N)$ distribution for $N\ge 3$ is 
also uniformly distributed. However, if the phases of $q_{mn}$ and 
$q_{nm}^{*}$ are uncorrelated, 
then $V [Tr\{QX_{Z(2)}\}]\approx V [Tr\{QX_{Z(N)}\}]$, ($N\ge 3$)
which, we will see, is apparently the case for the operators 
studied here.

\section{Partitioning the Problem}

\subsection{Basic Idea}

By \lq\lq partitioning" I mean replacing the single noise vector
problem,
\begin{eqnarray}
Mx = \eta,
\end{eqnarray}
which yields a complete output column, $\sum_{k} 
M^{-1}_{ik}\eta_{k}$, from
a single input noise vector, $\eta$, with a problem
\begin{eqnarray}
Mx^{p} = \eta^{p}, p=1,\dots P,
\end{eqnarray}
where the $\eta^{p}$ have many zeros corresponding to
some partitioning scheme. In this latter 
case it takes $P$ inverses to produce a complete measurement 
or sampling of a column of $M^{-1}$. 

For the unpartitioned problem (for $Z(N), N \ge 3$, say) 
\begin{eqnarray}
V [Tr\{QX \}] &=& \frac{1}{L}\sum_{m\ne n} |q_{mn}|^{2},\\ \nonumber
&\equiv & \frac{1}{L}N(N-1)<|q|^{2}>,
\end{eqnarray}
where I have defined the average absolute squared off diagonal
matrix element, $<|q|^{2}>$. For the partitioned 
problem the total variance includes
a sum on $p$,
\begin{eqnarray}
\sum_{p=1}^{P} V [Tr\{QX_{p}\}] &=& \frac{1}{L}\sum_{p=1}^{P}
\sum_{m_{p}\ne n_{p}}|q_{m_{p}n_{p}}|^{2}, \\ \nonumber
&\equiv & \frac{1}{L}N(\frac{N}{P}-1)<|q_{P}|^{2}>.
\end{eqnarray}

In order for this method to pay off in terms of computer time, 
one needs that
\begin{eqnarray}
\sum_{p=1}^{P} V [Tr\{QX_{p}\}] \le \frac{1}{P} V [Tr\{QX \}], \\
\Rightarrow <|q_{P}|^{2}> \le 
\left(\frac{N-1}{N-P}\right)
<|q|^{2}>. 
\end{eqnarray}
The goal of partitioning is to
avoid some of the large off-diagonal
matrix elements so that in spite
of doing $P$ times as many inverses, a smaller variance is
produced for the same amount of computer time.
The spaces partitioned can be space-time, color or
Dirac or some combination. I have found that
partitioning in Dirac and color spaces can 
strongly affect the results.

\subsection{Simulation Description}

I consider all local operators,
${\bar \psi }(x)\Gamma\psi (x)$, as well as point-split 
versions of the
vector and axial vector operators. This means 
16 local operators
and 8 point-split ones, making a total of 24,
which are listed below. For each operator there are
both real and imaginary parts, but in each case 
one may show via the quark propagator identity
$S = \gamma_{5}S^{\dagger}\gamma_{5},
\label{identity}$
that only the real {\it or} the imaginary part of 
each local or nonlocal
operator is nonzero on a given configuration
for each space-time point. 
However, this identity
is not respected exactly by noise methods, so the 
cancellations are actually only
approximate configuration by configuration.
However, the knowledge that one part is purely noise
allows one to simply drop that part in the
calculations, thus reducing the variance
without biasing the answer.

The operators I consider are:

\vspace{.3cm}
\noindent
$\bullet$ Scalar:\quad  $Re[{\bar\psi}(x)\psi(x)]$ \\
$\bullet$ Local Vector:\quad $Im[{\bar\psi}(x)\gamma_{\mu}\psi(x)]$ \\
$\bullet$ Point-Split Vector: 
  
$\kappa Im[{\bar\psi}(x+a_{\mu})(1+\gamma_{\mu})U^{\dagger}_{\mu}(x)\psi(x)-
{\bar\psi}(x)(1-\gamma_{\mu})U_{\mu}(x)\psi(x+a_{\mu})]$ \\
$\bullet$ Pseudoscalar: \quad $Re[{\bar\psi}(x)\gamma_{5}\psi(x)]$ \\ 
$\bullet$ Local Axial: \quad  
$Re[{\bar\psi}(x)\gamma_{5}\gamma_{\mu}\psi(x)]$ \\ 
$\bullet$ Point-Split Axial:
 
$\kappa Re[{\bar\psi}(x+a_{\mu})\gamma_{5}\gamma_{\mu}
U^{\dagger}_{\mu}(x)\psi(x)+
{\bar\psi}(x)\gamma_{5}\gamma_{\mu}U_{\mu}(x)\psi(x+a_{\mu})]$ \\
$\bullet$ Tensor: \quad 
$Im[{\bar\psi}(x)\sigma_{\mu\nu}\psi(x)]$
\vspace{.3cm}

I actually consider the zero momentum version of these operators,
summed over both space and time.

The sample noise variance in $M$ 
quantities $x_{i}$ is given by
the standard expression:
\begin{equation}
V_{noise} = \frac{1}{M-1}\sum_{i=1,M}(x_{i}-{\bar x})^{2}
\label{variance}
\end{equation}
What I concentrate on here are the relative 
variances between the different methods. 
Since the squared noise error in a single 
configuration is given by
\begin{equation}
\sigma_{noise}^{2}=\frac{V_{noise}}{M},
\label{error}
\end{equation}
the ratio of variances gives a direct measure 
of the multiplicative ratio of noises, and thus the computer time,
necessary to achieve the same noise error. However, the 
variance itself does not take into account
the extra $P$ inverses done when problem is partitioned.
In order to measure the relative efficiency of different
partitionings, I form what I call pseudo-efficiencies ratios (\lq\lq PE"), 
which are defined by
\begin{equation}
{\rm PE(\frac{method1}{method2})}\equiv
\frac{P_{method1}(V_{noise})^{method1}}
{P_{method1}(V_{noise})^{method2}},
\label{Peff}
\end{equation}
where $P_{method}$ are the number of partitions required
by the method. I refer to these ratios as \lq\lq pseudo" efficiencies
since I do a fixed number of iterations
for all of the operators I consider. It could very well be 
that different methods will require significantly different 
numbers of iterations of conjugate-gradient
or minimum residual for the same level of accuracy.
One desires to find the lowest PE ratio for a
given operator. \footnote{Of course an evaluation
via $N$ partitioning on an $N\times N$ matrix
yields a PE numerator factor of zero relative to other
methods since the variance is exactly zero in this case. 
This is of course prohibitively expensive, but
it suggests that more efficient partitionings are 
possible for large computer budgets. 
Thanks to M. Peardon for bringing this point out.}

I display PE ratio results for Wilson fermions in a
$16^{3}\times 24$, $\beta=6.0$ lattice with $\kappa=0.148$
in Table 1, which follows on the next page. 
(Part of this Table also appeared in
Ref.\cite{me1}.) I will examine 3 partitionings:
\vspace{.3cm}
\noindent

{$\bullet$ $Z(2)$ unpartitioned ($\lq\lq P=1\ Z(2)"$);

{$\bullet$ $Z(2)$ Dirac partitioned ($\lq\lq P=4\ Z(2)"$);

{$\bullet$ $Z(2)$ Dirac and color partitioned ($\lq\lq P=12\ Z(2)"$).
\vspace{.3cm}

My Dirac gamma matrix representation is:
\begin{equation}
\gamma_{i}=\left( \begin{array}{cc} 0 &\sigma_{i}\\ \sigma_{i} &0 
\end{array} \right) , 
\gamma_{4}=\left( \begin{array}{cc} 1 &0 \\ 0 & -1 
\end{array} \right) ,
\gamma_{5}=\gamma_{1}\gamma_{2}\gamma_{3}\gamma_{4}=
\left( \begin{array}{cc} 0 &-i \\ i & 0 
\end{array} \right) .
\end{equation}

In Table 1 I list
the relative PEs of the two partitioned methods relative to
the unpartitioned case. Referring to the above list of 
the real or imaginary parts of operators
my notation here is as follows: \lq\lq
Scalar" stands for the operator ${\bar \psi}\psi$, \lq\lq
Local Vector 1" for example stands for the operator 
${\bar \psi}\gamma_{1}\psi$,\lq\lq
P-S Vector 1" stands for the 1 component of
the point split vector current,\lq\lq pseudoscalar" stands for 
${\bar \psi}\gamma_{5}\psi$,\lq\lq
Local Axial 1" stands for the operator 
${\bar \psi}\gamma_{5}\gamma_{1}\psi$,\lq\lq
P-S Axial 1" stands for the point split axial 1 component, and
\lq\lq Tensor 41" stands for example for the operator
${\bar \psi}\sigma_{41}\psi$.

There are extremely large variations in the behaviors of the
operators listed in Table 1 under identical partitionings.
Of the partitionings considered it is most
efficient to calculate scalar and vector operators 
with an unpartitioned simulation. 
On the other hand, it is far more
efficient to calculate the pseudoscalar in a Dirac
and color partitioned manner.
Notice the entries for the 1,2 components of the
axial current (both local and point split) do not 
behave like the 3,4 components under pure Dirac
partitioning, but they do when Dirac and color partitionings
are combined. Four of the tensor operators respond best
to a pure Dirac partitioning, while the other two prefer
a partitioning in Dirac and color spaces combined.
The ratio of the largest to the smallest entry in
the right hand column is about 800! 

As pointed out in Section 2, the variance of $Z(2)$ 
and $Z(N)$ ($N\ge 3$)
noises are in general different. For this reason 
I also investigated 
partitioning using $Z(4)$ as well as volume 
(gauge variant) noises, but
there do not seem to be large factors to be gained relative
to the $Z(2)$ case. The SESAM collaboration also
has seen the efficacy of partitioning (in Dirac space) for axial
operators\cite{sesam}.

\begin{table}
\caption{The pseudoefficiency (PE) ratios associated 
with the methods indicated.}
\centering
\begin{tabular}{ccc}
\hline
{ Operator\quad} & 
{ PE($\frac{\rm{P=4\ Z(2)}}
{\rm{P=1\ Z(2)}})$} &
{ PE($\frac{\rm{P=12\ Z(2)}}{\rm{P=1\ Z(2)}}$)} \\
\hline
  
&  &  \\
\quad\quad{\bf Scalar}\quad\quad\quad  
&  $2.83\, \pm\, 0.47$ & $10.9\, \pm\, 2.5$ \\   
{\bf Local Vector\,1} &  $2.38\, \pm\, 0.65$ & $8.71\, \pm\, 1.8$ \\  
{\bf Local Vector\,2} &  $2.50\, \pm\, 0.53$ & $12.1\, \pm\, 2.9$ \\ 
{\bf Local Vector\,3} &  $3.60\, \pm\, 1.00$ & $11.4\, \pm\, 2.4$ \\ 
{\bf Local Vector\,4} &  $3.41\, \pm\, 0.60$ & $16.3\, \pm\, 3.4$ \\
 
{\bf P-S Vector\,1}   &  $2.63\, \pm\, 0.56$ & $9.94\, \pm\, 2.2$ \\ 
{\bf P-S Vector\,2}   &  $2.27\, \pm\, 0.44$ & $11.0\, \pm\, 2.3$ \\
{\bf P-S Vector\,3}   &  $3.52\, \pm\, 0.74$ & $11.4\, \pm\, 1.5$ \\
{\bf P-S Vector\,4}   &  $3.87\, \pm\, 0.49$ & $15.5\, \pm\, 4.2$ \\

{\bf Pseudoscalar}  &  $0.698\, \pm\, 0.15$ & $0.0201\, \pm\, 0.0043$ \\

{\bf Local Axial\,1}  &  $0.114\, \pm\, 0.021$ & $0.144\, \pm\, 0.029$ \\
{\bf Local Axial\,2}  &  $0.126\, \pm\, 0.020$ & $0.146\, \pm\, 0.037$ \\
{\bf Local Axial\,3}  &  $1.13\, \pm\, 0.19$ & $0.162\, \pm\, 0.038$ \\
{\bf Local Axial\,4}  &  $2.26\, \pm\, 0.24$ & $0.187\, \pm\, 0.035$ \\

{\bf P-S Axial\,1}  &  $0.167\, \pm\, 0.040$ & $0.151\, \pm\, 0.018$ \\
{\bf P-S Axial\,2}  &  $0.110\, \pm\, 0.032$ & $0.114\, \pm\, 0.028$ \\
{\bf P-S Axial\,3}  &  $1.67\, \pm\, 0.35$ & $0.186\, \pm\, 0.049$ \\
{\bf P-S Axial\,4}  &  $1.85\, \pm\, 0.21$ & $0.238\, \pm\, 0.036$ \\

{\bf Tensor\,41} &  $1.07\,   \pm\, 0.30$  &   $0.295\, \pm\, 0.049$ \\
{\bf Tensor\,42} &  $0.345\,  \pm\, 0.076$ &   $0.0889\, \pm\, 0.011$ \\
{\bf Tensor\,43} &  $1.32\,   \pm\, 0.43$  &   $0.398\, \pm\, 0.13$ \\
{\bf Tensor\,12} &  $1.12\,   \pm\, 0.25$  &   $0.376\, \pm\, 0.066$ \\
{\bf Tensor\,13} &  $0.116\,  \pm\, 0.024$  &   $0.363\, \pm\, 0.053$ \\
{\bf Tensor\,23} &  $0.0314\, \pm\, 0.0058$ &   $0.0751\, \pm\, 0.016$ \\ 
\hline

\end{tabular}
\end{table}

\section{Perturbative Noise Subtraction}

\subsection{Description of Algorithm}

Consider ${\tilde Q}$ such that
\begin{equation}
<Tr\{{\tilde Q}X \}>=0.
\end{equation}
One can then form
\begin{equation}
<Tr\{(Q-{\tilde Q})X \}>=<Tr\{ QX \}>.
\end{equation}
However,
\begin{equation}
V[Tr\{(Q-{\tilde Q})X \}]\ne V[Tr\{Q X \}].
\end{equation}

As we have seen for $Z(N)$ ($N\ge 2$), the variance 
comes exclusively from off
diagonal entries. So, the trick is to try to find 
matrices ${\tilde Q}$ which are
traceless (so they do not affect the expectation value) 
but which mimic the off-diagonal part of $Q$ as much as 
possible to reduce the variance.

The natural choice is simply to choose as 
${\tilde Q}$ the {\it perturbative}
expansion of the quark matrix. Formally, one has
($I,J=\{x,a,\alpha \}$)
\begin{equation}
(M^{-1})_{IJ}=\frac{1}{\delta_{IJ}-\kappa P_{IJ}},
\end{equation}
where
\begin{eqnarray}
P_{IJ}=\sum_{\mu}[(1+\gamma_{\mu})U_{\mu}(x)
\delta_{x,y-a_{\mu}}+
(1-\gamma_{\mu})U_{\mu}^{\dagger}(x-a_{\mu})
\delta_{x,y+a_{\mu}}].
\end{eqnarray}

Expanding this in $\kappa$ gives the perturbative 
(or hopping parameter) expansion,

\begin{eqnarray}
M^{-1}_{p} =  &I&  + \kappa P  + \kappa^{2} P^{2} +
\kappa^{3} P^{3} + \cdots .
\end{eqnarray}

One constructs $<\eta_{j}
(M^{-1}_{p})_{ik}\eta_{k}>$ and subtracts it from $<\eta_{j}
M^{-1}_{ik}\eta_{k}>$, where
$\eta$ is the noise vector. Constructing 
$<\eta_{j} (M^{-1}_{p})_{ik}\eta_{k}>$ is
an iterative process and is easy to code 
and extend to higher powers on the
computer. I will iterate up to 10th order in $\kappa$.

One can insert coefficients in front of the 
various terms and vary them to find
the minimum in the variance, but such coefficients 
are seen to take on their perturbative
value, at least for high order expansions\cite{liu}. 
However, see also Ref.\cite{nilmani} where in 
low orders there is apparently an advantage to this procedure.
Interestingly, significant subtraction improvements 
occur in some operators even in 0th order (point
split vectors and two tensor operators.)

For a given operator,
${\cal O}$, the matrix ${\cal O}M^{-1}_{p}$ 
encountered in the context of $<{\bar \psi}{\cal
O}\psi>=-Tr({\cal O}M^{-1})$ is not traceless. 
In other words, one must re-add
the perturbative trace, subtracted earlier, 
to get the full, unbiased answer.
How does one calculate the perturbative part?
The exact way is of course to
explicitly construct all the gauge invariant paths 
(up to a given $\kappa$ order) for a given operator.
Another approach is to subject the perturbative
contribution to a separate Monte Carlo estimation.
This is the approach taken here. 
Local operators require perturbative corrections starting at 4th 
order (except a trivial correction for ${\bar \psi}\psi$ at
zeroth order) and point split ones require corrections
starting instead at 3rd order. Because one
is removing the bias (perturbative trace) 
by a statistical method, I refer to this as a 
\lq\lq statistically unbiased" method. Some
efficiency considerations in carrying out this 
procedure will be discussed in Section 5. Other
versions of subtraction methods in the context
of lattice evaluations of disconnected diagrams
may be found in Refs.\cite{michael} and \cite{sesam2}.

\subsection{Numerical Results}

I am carrying out this numerical 
investigation in an unpartitioned 
sense ($P=1$). The operators which respond best to this 
partitioning, as discussed previously, are the scalar
and local and point-split vector currents, and attention
will be limited here to these cases. The effect of
combining the partition and subtraction methods has
not yet been investigated.
I show the ratio of unsubtracted variance divided by 
subtracted variance, $\bf V_{\rm
unsub}/V_{\rm sub}$ in Figs. 1 and 2. Factors larger 
than one give the multiplicative 
gain in computer time one is achieving.
The lattices are again Wilson $16^{3}\times 24$, $\beta=6.0$.

Notice the approximate linear rise in the variance ratio
as a function of subtraction order 
for the point-split vector charge density at
both $\kappa=0.148$ and $0.152$, Figs. 1 and 2 respectively.
Also notice that even at $S=0$ (subtracting the Kronecker delta)
there is a reduction in the variance.
The slope of the subtraction graph at $\kappa=0.148$ is
about 3.5; the slope at $\kappa=0.152$ is reduced to a
little under 3.0. Although I do not show the results here, 
the same linear behavior
is evident in ${\bar \psi}\psi$ and the local vector operators
although their slopes are considerably smaller.

My final results are summarized in Fig. 3, which gives the reduction
in the variance in the scalar and vector operators after
a 10th order subtraction has been made at $\kappa=0.148$. It is
not known why the point split vector current responds 
the best to subtraction. The 10th order point-split vector, 
local vector, and scalar variance ratios change 
from $\sim 35$, $\sim 12$, and $\sim 10$ at 
$\kappa =0.148$, to $\sim 25$ $\sim 10$, and $\sim 5$ at 
$\kappa =0.152$, respectively.
These are all zero momentum operators.
Although the results are not shown, 
I have found essentially identical
results to the above for momentum transformed data,
necessary for disconnected form factors.
Perturbative subtraction methods will thus be extremely useful
in lattice evaluations of nucleon strangeness form factors
using the point split (conserved) form of the vector current.

\begin{figure}
\vspace{8.0cm}
\special{illustration 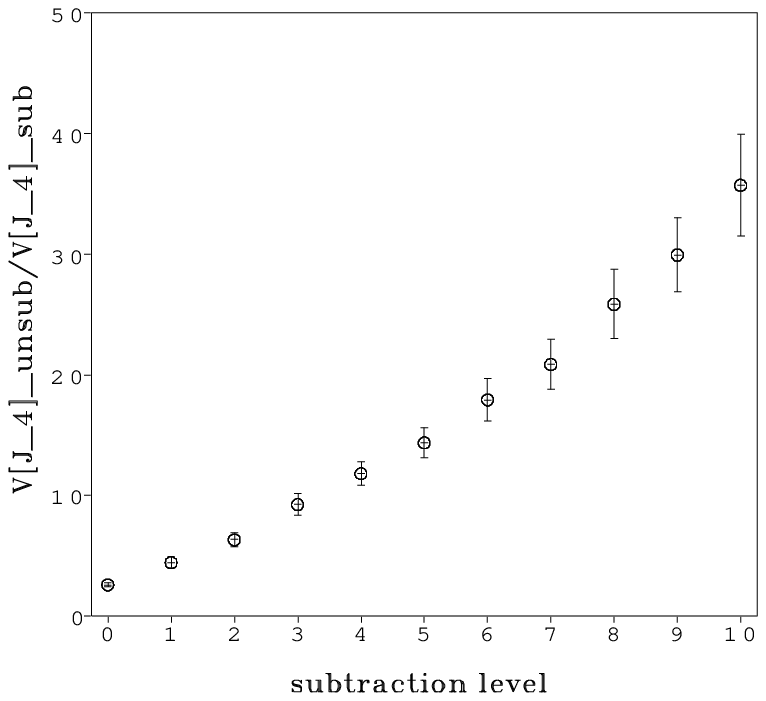}
\caption{Effect of the level of perturbative subtraction, up to 
tenth order in $\kappa$, on the ratio of unsubtracted 
divided by subtracted noise variance for the zero 
momentum point-split (conserved) charge density operator, 
$J_{4}$, at $\kappa=0.148$.}
\end{figure}

\begin{figure}
\vspace{8.5cm}
\special{illustration 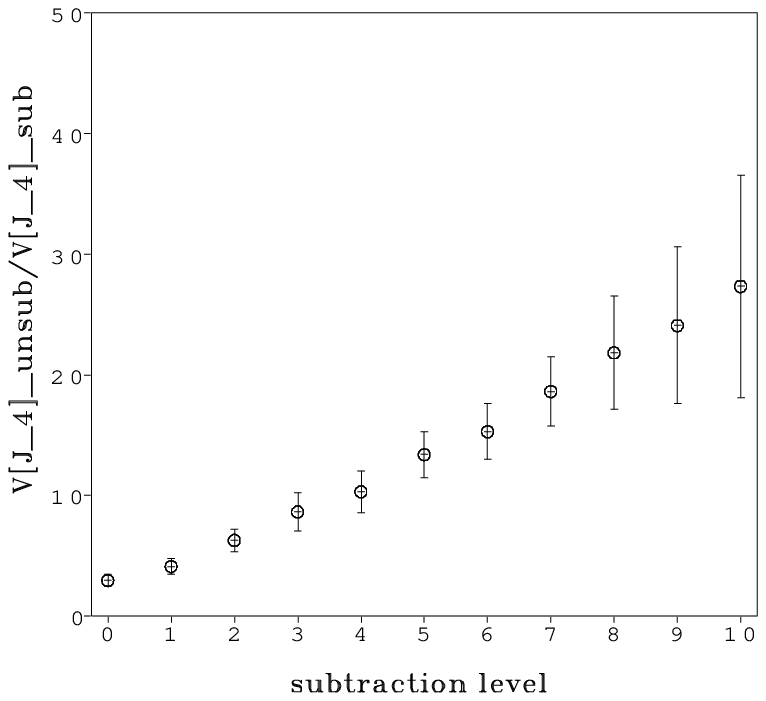}
\caption{Same as Fig. 1 but for $\kappa=0.152$.}
\end{figure}

\begin{figure}
\vspace{7.5cm}
\special{illustration 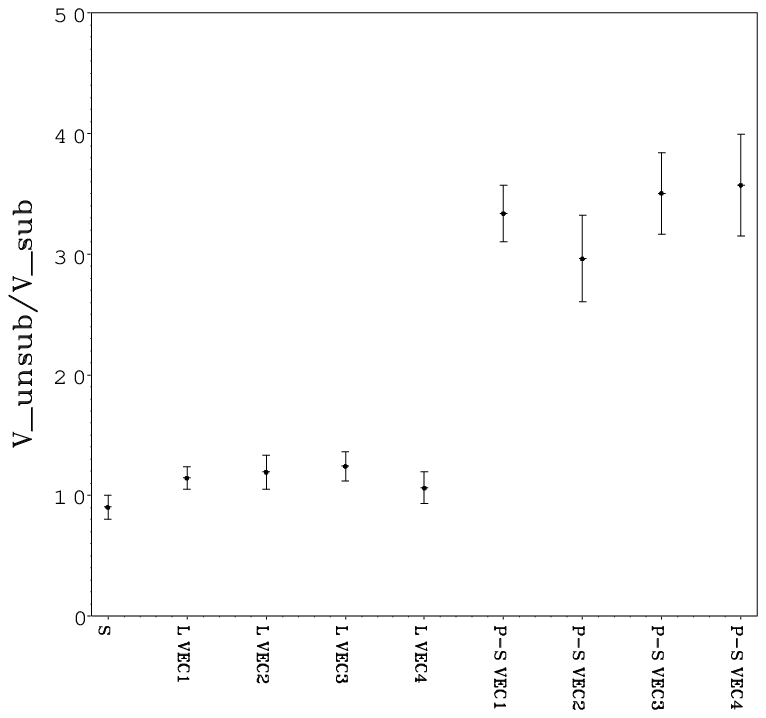}
\caption{Graphical presentation of the unsubtracted 
variance divided by subtracted variance of 9 different
lattice operators after tenth order subtraction at 
$\kappa=0.148$ for Wilson fermions. (Abbreviations used for operators: 
S=scalar; L VEC = local vector; P-S VEC = point split vector).}
\end{figure}

\section{Efficiency Considerations}

\subsection{Fixed Noise Case}

Let me close with some simple observations regarding
statistical errors in flavor singlet Monte Carlo simulations. 
There are two sources of variance in such simulations:
gauge configuration and noise. Given $N$ configurations
and $M$ noises per configuration, the final error bar 
on a given operator is given by
\begin{equation}
\sigma = \sqrt{\frac{V_{noise}}{NM}+\frac{V_{gauge}}{N}},
\label{error}
\end{equation}
where $V_{guage}$ and $V_{noise}$ are the gauge configuration
and noise variances. For fixed $NM$ (total number of noises), it
is clear that Eq.(\ref{error}) is minimized by taking $M=1$.
Thus, in this situation it is best to use a single noise
per configuration. This simple result is modified by real world considerations
of overheads. For example, if one assumes that there is an overhead
associated with generating configurations and fixes instead
the total amount of computer time,
\begin{equation}
T = NM + G_{N}N,
\end{equation}
where $G_{N}$ is the appropriately scaled configuration generation
time overhead, then one finds instead that
\begin{equation}
M = \frac{S_{noise}}{S_{gauge}}\sqrt{G_{N}},
\end{equation}
is the best choice. Note that the ratio 
$S_{noise}/S_{gauge}$ can have a wide range of values
for various operators, and one is no longer guaranteed that
$M=1$ is optimal.

\subsection{Fixed Configuration Case}

Another common real world situation is where $N$, the number of 
configurations, is fixed.
In the context of the perturbative subtraction algorithm, 
one should now maximize the number of effective noises for a 
given computer budget. The effective number of noises is
\begin{equation}
M_{eff} = M(X+S\Delta s),
\label{eff}
\end{equation}
where $M_{eff}$ replaces $M$ in Eq.(\ref{error}). ($M$ retains its
meaning as the actual number of gauge
field noises.) $S$ is the subtraction
order, ranging from 0 to 10 in Figs. 1 and 2, and $\Delta s$
is the slope. $X$ is the factor one obtains from this method
at $S=0$, without extra overhead. (One sees in Figs. 1 and 2 that
this factor is about 2 for the point split charge operator.)
I am assuming that the reduction 
in the variance is approximately linear in subtraction order
$S$. $S$ is imagined to be a continuously variable quantity.
The total time per gauge field is
\begin{equation}
T_{N} = (MT_{M}+ST_{S}),
\end{equation}
which is kept fixed as $M_{eff}$ is varied. 
$T_{M}$ is the noise time overhead 
and $T_{S}$ is the subtraction time overhead. 
The optimum choices for $S$ and $M$ are now
\begin{eqnarray}
S = \frac{T_{N}}{2T_{S}} - \frac{X}{2\Delta s}, \\
M = \frac{T_{N}}{2T_{M}} + \frac{XT_{S}}{2\Delta sT_{M}},
\end{eqnarray}
resulting in
\begin{equation}
M_{eff} = \frac{\Delta sT^{2}_{N}}{4T_{S}T_{M}} + \frac{T_{N}X}{2T_{M}} 
- \frac{X^{2}T_{S}}{4\Delta sT_{M}}.
\end{equation}
The interesting aspect of this last result is that
the effective number of noises, $M_{eff}$, is now
quadratic in $T_{N}$. This is a consequence of our observation that
the slopes in Figs. 1 and 2 are approximately linear in $S$.
The immediate implication is that for large $T_{N}$
the noise error bar can be made to vanish like the inverse
of the simulation time rather than as the usual 
inverse square root, at least
in the range of the existing linear behavior.

These equations are also helpful when one has an exact 
analytical representation of the trace of the perturbative
series up to some order, $S_{exact}$, making $T_{S}$
zero. Then, by comparing the $M_{eff}$ values in the two cases,
one may find the lowest value of $S$, $S_{lowest}$, such that 
$M_{eff}^{lowest}\ge M_{eff}^{exact}$. For example,
when $X\approx 0$ and for common values of $T_{N}$ and $T_{M}$
in the two simulations, one has
\begin{equation}
S_{lowest}= 2S_{exact}.
\end{equation}
That is, for the extra Monte Carlo overhead
to pay off, one must attempt to subtract to at least twice as 
high an order in $\kappa$ as the exact evaluation. Since subtraction
is always exact for nonlocal operators up to second order
and for local ones up to third order, making $S_{lowest}\ge 6$
will usually result in a more efficient simulation than
the default exact one. 

\section{Summary}

We have seen that significant
savings in computer resources may be obtained by 
partitioning the Wilson matrix appropriately.
An efficient partitioning reduces the variance 
of an operator by leaving out the largest off-diagonal
matrix elements of the quark propagator so that in spite
of having to do more inversions, a smaller variance is
produced for the same amount of computer time.
A numerical investigation in Dirac and color spaces
revealed efficient partitionings for
24 local and nonlocal operators summarized near the end of Section 3. 

We have also seen that large time 
savings are possible using subtraction methods
for selected operators in the context
of unpartitioned noise simulations. This method was shown 
to be effective for the scalar and local vector currents, 
but most effective for the point-split vector currents. 
Since perturbative subtraction is based on the hopping parameter
expansion of the quark propagator, such methods can 
become less effective at 
lower quark masses, although we found the
variance reduction was still quite significant
for the point-split vector operator at $\kappa=0.152$. 
Similar methods can be devised for other operators (axial,
pseudoscalar, tensor) by implementing these ideas 
in the context of Dirac/color partitioned noise methods.

There are still a number of open questions here. 
The reasons for the strange partitioning patterns found
in Section 3 are not known. In addition, the reason why the variance
of some operators respond more sensitively to
perturbative subtraction than others is obscure. 
These questions are important because their answers could
lead one to better simulation methods.
Another question is how far
the linear slopes in Figs. 1 and 2 persist at
high subtraction orders. Since the perturbative expansion
of the Wilson matrix does not converge at small quark mass,
the slope of such curves probably levels off at high 
enough $S$. We have seen, however, that before this leveling off
occurs the number of effective noises 
grows quadratically in the simulation time.
It was pointed out that this implies that
the noise error bar can be made to vanish like the inverse
of the simulation time in the range of the existing linear behavior.

\section{Acknowledgements}

This work was supported in part by National Science Foundation Grant
No. 9722073, and generous grants of computer time
on the SGI Origin 200 computer from
the National Center for Supercomputing Applications (NCSA)
in Urbana/Champaign, Illinois. I thank the University 
of Kentucky Department of Physics and
Astronomy and the Special Research
Centre for the Subatomic Structure of Matter,
University of Adelaide for their hospitality
and invitations while portions of this work were underway. I also
thank the conference organizers at the
University of W\"uppertal for their invitation to this workshop.


\end{document}